\documentclass[12pt]{article}
\usepackage{amsfonts}
\usepackage[english]{babel}
\input epsf
\usepackage{amssymb}
\usepackage[dvips]{graphicx}
\usepackage{amscd}
\usepackage{mathrsfs}
\usepackage{amsmath}
\usepackage{amscd}
\def\tr{{\rm tr}}

\newcommand{\be}{\begin{equation}}
\newcommand{\ee}{\end{equation}}
\newcommand{\bea}{\begin{eqnarray}}
\newcommand{\eea}{\end{eqnarray}}

\begin{document}
\linespread{1} \flushbottom
%



\begin{center}
{\Large\bf{Construction of the Noncommutative Complex Ball}}
\vskip1cm
Zhituo Wang\\ {\it Dipartimento di Matematica, Universit\`a di Roma Tre\\
Largo S. L. Murialdo 1, 00146 Roma, Italy}\\
email: zhituo@mat.uniroma3.it
\end{center}

\begin{abstract}
We describe the construction of the noncommutative complex ball whose commutative analog is the
Hermitian symmetric space $D=SU(m,1)/U(m)$, with the method of coherent state quantization. In the commutative limit we obtain the standard manifold. We consider also a quantum field
theory model on the noncommutative manifold.

\end{abstract}

\begin{flushleft}
Mathematics Subject Classification (2010): 22E70, 53D55
\end{flushleft}

\section{Introduction}
It is believed that the ordinary spacetime becomes fuzzy when we are approaching the Planck scale hence the
traditional differential geometry for describing the space-time should be replaced by noncommutative geometry \cite{connes}.
The simplest noncommutative manifold is the Moyal space, a symplectic manifold
generated by the noncommutative coordinates $x_\mu$, such that  $[x_\mu, x_\mu]=i\theta_{\mu\nu}$, where 
$\theta_{\mu\nu}=-\theta_{\nu\mu}$ 
are constants and independent of the coordinates. 

Quantum field theories defined on noncommutative space time (NCQFT) [2, 3, 4] are considered
as one possible way to explore the effects of quantum gravity.
The first well defined quantum field theory on 4 dimensional Moyal 
space is the Grosse-Wulkenhaar model \cite{GW1}. It is perturbative renormalisable to all orders 
and the beta function is vanishing \cite{beta1, betaall}. Hence it is also possibly constructively renormalizable\cite{rivass} or solvable \cite{gwsolve}. Recently the two dimensional Grosse-Wulkenhaar model has been constructed in \cite{Wanggw}.

Since noncommutative quantum fields theories are better behaved than their
commutative counterparts, it is very natural to construct more nontrivial noncommutative manifolds and consider
physics models over them. The main idea of the construction consists in reformulating the 
geometry of a manifold in terms of an algebra of smooth functions defined on it, and then to generalize the
differential calculus to a noncommutative algebra. In this paper we construct the noncommutative manifold $\hat D$ whose commutative analog is the complex ball defined as the coset space $D=SU(m,1)/S(U(m)\times U(1))$, 
with the method of coherent state quantization. 

We concentrate only on the case $m=2$, since in this case the manifold to be studied has real dimension $4$, which is physically more interesting. The generalizations to larger $m$ follow very easily. 

The construction of the noncommutative coset space SU(2,1)/U(2) has been
also studied by [14], [15] with the method of Berezin-Toeplitz quantization and by
by [16] with the method of ”WKB quantization”. The interested reader could go to
the references for details. The construction of the noncommutative matrix ball $B=SU(2,2)/S(U(2)\times U(2))$
has been and the corresponding noncommutative quantum field theory model has also been studied \cite{GPW}.

This paper is organized as follows. In Section 2 we give an general
introduction to the representations of the Group $SU(m,1)$ and the interesting properties of the complex ball $D$. In section 3, we present a simple harmonic oscillators realization of the most
degenerate discrete series of representations of $SU(m,1)$ group, which will be used to construct the coherent states in section 4
. The generalization to an arbitrary $SU(m,n)$ group is given in Appendix B.

In Section 4 we construct the system of coherent states \cite{Perelomov} for the
representation in question and construct the noncommutative manifold $\hat D$, which is the noncommutative complex ball we want to construct. Here we give a closed formula for the star-product of functions defined on  $\hat D$.

In Section 5 we consider a quantum theory of complex scalar fields
on $\hat D$ and calculate the amplitude of the tadpole graph. We find that this amplitude is finite.

The construction of $\hat D$ has been studied also
by \cite{Upmeier}, \cite{Jakim} with the method of Berezin-Toeplitz quantization and by
by \cite{BGR} with the method of "WKB quantization". The interested
reader could go to the references for details.

\section{The $SU(m,1)$ group and its Lie algebras}
The group $G= SU(m,1)$, $m\ge 1$ is the subgroup of 
$SL(m+1,C)$ matrices satisfying:
\begin{equation}\label{def}
g=\left(\begin{array}{cc}a& b\\ c& d\end{array}\right)\in SU(m,1)\Rightarrow
g^\dagger \Gamma g=g \Gamma g^\dagger=\Gamma ,\ \ \
\Gamma=\left(\begin{array}{cc}{I_{m\times m}}&0\\ 0& -1\end{array}\right).
\end{equation}
Here $a$ is an $m\times m$ dimensional matrix, $b$ is an $m\times 1$ matrix, $c$ is a
$1\times m$ matrix and $d$ is a complex number. $I_{m\times m}$ means the  $m\times m$ dimensional unit matrices and $0$ means the blocks of zeros.

From equation (\ref{def}) we have two equivalent sets of constraints:
\begin{equation}
a^\dagger a = I+c^\dagger c,\ \ \ d^\dagger d=|d|^2 = 1+b^\dagger b,\
a^\dagger b=c^\dagger d
\end{equation}
or
\begin{equation}
a a^\dagger= I+b b^\dagger,\ d d^\dagger =|d|^2= 1+c c^\dagger ,\ a
c^\dagger = b d^\dagger,
\end{equation}
where $\dagger$means Hermitian conjugation for matrices and complex conjugate for complex numbers.
We could easily find that both $a$ and $d$ are invertible.

The maximal compact subgroup $K = S(U(m)\times U(1))$ of $G$ is
defined by the matrices
\begin{equation}\label{compact}
k=\left(\begin{array}{cc}k_1&0\\ 0& k_2\end{array}\right)\in K,\ \
k_1, k_2\ -\ \mbox{unitary},\ \ \ \det(k_1\ k_2) = 1. \end{equation} \vskip0.5cm


Let $\textbf{g} = su(m,1)$ be the Lie algebra of $SU(m,1)$, so it is real and semi-simple. $g$ is generated by the $(m+1)\times(m+1)$ matrices $X$ satisfying:
\begin{equation}\label{lie} X^\dagger \Gamma\ +\ \Gamma X\ =\ 0,
\end{equation}
So we can easily find that every element $X\in \textbf{g}$ has the form:
\begin{equation} X\ =\ \begin{pmatrix}
A & B \\
B^\dagger & D \\
\end{pmatrix},
\end{equation}
where $A$ is an $m\times m$ dimensional matrix, $B$ is an $m\times 1$ dimensional matrix and D is a complex number,
such that
\begin{equation}
 A^\dagger = -A,\ D^\dagger = -D,\ \
\mbox{tr}(A)+\mbox{tr}(D)\,=\,0,\nonumber
\end{equation}

The Cartan decomposition of $\textbf{g}$ reads (see \cite{Kirillov}\cite{knapp}):
\be \textbf{g}=\textbf{k}+\textbf{p},\ee
where $\textbf{k}=\{\begin{pmatrix}
A & 0 \\
0 & D \\
\end{pmatrix}\}$ is the Lie algebra of the maximal compact subgroup of G and .
$\textbf{p}=\{\begin{pmatrix}
0 & B \\
B^\dagger & 0 \\
\end{pmatrix}\}$ correspond to the noncompact part of $\textbf{g}$. It  is a linear space but not a Lie algebra.


%
%

Let $\textbf{a} \in \textbf{p}$ be a maximal Abelian subalgebra. We
could choose for $\textbf{a}$ the set of all matrices of the form
\begin{equation}
H_t=\begin{pmatrix}
O_{(m-1)\times(m-1)} & O_{(m-1)\times 1} & O_{(m-1)\times 1} \\
O_{1\times(m-1)} & 0 & t \\
O_{1\times(m-1)} & t & 0 \\
\end{pmatrix}
\end{equation}
where $t$ is a real number.

Define the linear functional over $H_t$ by $\alpha(H_t)=t$, the
roots of $(\textbf{g},\textbf{a})$ are given by
\begin{equation}
\pm \alpha,\  \pm 2\alpha,
\end{equation}
with multiplicities $m_{\alpha}=2$ and $m_{2\alpha}=1$.

Define
\begin{equation}
\delta:= \{\textbf{a}_t|\  \textbf{a}_t=\exp H_t,\ H_t\in
\textbf{a}\}.
\end{equation}
so we have
\begin{equation}
\textbf{a}_t=\begin{pmatrix}
I & O & 0\\
O & \cosh t & \sinh t \\
0 & \sinh t& \cosh t \\
\end{pmatrix},
\end{equation}
where the symbol $I$ stands for the identity matrix and $O$ is the matrix with entries
$0$ .

On the root system we choose that the positive Weyl chamber  given
by $C^{+}=\{t\}$ with $t>0$. Then the positive roots are $\alpha$
and $2\alpha$. And the simple root is $\alpha$.


Now we consider the $K\delta K$ decomposition of an arbitrary group
element $g\in G$:
\begin{equation}\label{cartan}
g=k\delta q^\dagger,
\end{equation}
where $k,\  q\in S(U(m)\times U(1))$.

We could write the Haar measure of group $g$ as:
\begin{equation}
dg=dg(t, k, q)=\rho(t)dt dk dq,
\end{equation}

where $dk$ and $dq$ are normalized Haar measure on the maximal
compact subgroup $U(m)=S(U(m)\times U(1))$, and $\rho(t)dt$ is the
measure on the noncompact group element. The explicit form of
$\rho(t)$ could be derived from the positive roots, see
(\cite{Kirillov}):
\begin{equation}
\rho(t)=\prod_{\alpha\in\Sigma^+}|\sinh\alpha(t)|^{m_\alpha},
\end{equation}
where $m_\alpha$ is the multiplicity of the positive roots.

So we have:
\begin{equation}
\rho(t)=\sinh^2 t\sinh 2t.
\end{equation}


\subsection{The complex ball}
The complex ball $D$ is defined as the coset space 
\be
D=G/K,
\ee
where
$G=SU(m,1)$ and $K=S(U(m)\times U(1))$ \cite{rudin}.

More precisely, for each element $g\in SU(m,1)$ we have the following Cartan decomposition:
\begin{equation}
  g=\left(\begin{array}{cc}N_1&Z N_2\\Z^\dagger N_1&N_2\end{array}\right) \left(\begin{array}{cc}K_1&0\\0&K_2\end{array}\right)
\end{equation}
where $\left(\begin{array}{cc}K_1&0\\0&K_2\end{array}\right)\in S(U(m)\times U(1))$ is an element of the maximal subgroup,
$N_1=(I-ZZ^\dagger)^{-1/2}$, $N_2=(1-Z^\dagger Z)^{-1/2}$, and
\begin{equation}\label{decomp}
Z=bd^{-1}=\left(\begin{array}{cc} z_1\\z_2\\ \cdots\\z_m \end{array}\right),\  Z^\dagger=(z_1^\dagger, z_2^\dagger,\cdots z_m^\dagger).
\end{equation}

So the complex ball $D$ can be represented by the complex vectors $Z$:
\begin{equation}
D=\{Z\,|\, 1-|Z|^2 > 0\}\ =\ \{Z\,|\, 1-|z_1|^2-|z_2|-\cdots
-|z_m|^2
> 0\}.
\end{equation}
The group action on D is given by:
\be
Z'=gZ=Z'=(aZ+b)(cZ+d)^{-1}.
\ee
$D$ is a pseudo-convex domain over which we could define the Hilbert spaces\cite{Upmeier} with the reproducing Bergman kernel:
\begin{equation}
K(W^\dagger, Z)=(1-W^\dagger Z)^{-N},
\end{equation}
where $N=m+1,m+2,\cdots$ is a
natural number characterizing the representation.

It is a topologically simply connected Hermitian symmetric space with K{\" a}hler structure \cite{Koba}.
The K{\"a}hler metric is defined by the derivations of the Bergman kernel:
\begin{equation}
 g_{i\bar j}\ =\frac{1}{N} \partial_{\bar{z}^i}\partial_{z^j}\,\log K(Z^\dagger,Z).
\end{equation}
More explicitly we have:
\begin{equation}
 g_{i \bar j}=[\frac{\delta_{ij}}{1-|Z|^2}+\frac{z_i\bar{z}_j
 }{(1-|Z|^2)^2}], \  \  g^{i \bar j}=(1-|Z|^2)(\delta_{ij}- \bar z_i z_j).
\end{equation}
Given a metric $g$ on a manifold $D$ we can calculate the Levi-Civita connection by:
\be
\Gamma^{i}_{ j k}=\frac{1}{2}g^{il}\ (\ \frac{\partial g_{lk}}{\partial x^j}+\frac{\partial g_{lj}}{\partial x^k}-\frac{\partial g_{jk}}{\partial x^l}\ ),
\ee 
where the coordinates $x^i$ meas either $z^j$ or $\bar z^j$, depending on the context and we will write
the indices of $\Gamma^{i}_{ j k}$ as $j$ or $\bar j$, respectively. 
It is well known from complex geometry that for a K\"ahler manifold only the Christoffel symbol with all holomorphic and anti-holomorphic indices are non-vanishing \cite{Koba}:
\begin{eqnarray}
\Gamma^{ l}_{ j k}\ =g^{ \bar s l } \partial_{z_j}
g_{k\bar s }\ =\frac{\delta_{kl}\bar z_j+\delta_{jl}\bar z_k}{1-|Z|^2},
\end{eqnarray}
and 
\begin{eqnarray}
\Gamma^{\bar l}_{\bar j\bar k}\ =g^{ s \bar l } \partial_{\bar z_j}
g_{s\bar k }\ =\frac{\delta_{kl}z_j+\delta_{jl}z_k}{1-|Z|^2}.
\end{eqnarray}
The only non-vanishing component of the Riemann tensor reads:
\begin{equation}
R_{i\bar j k\bar l}=g_{i\bar s}\frac{\partial \Gamma^{\bar s}_{\bar j\bar l}}{\partial z^k},
\end{equation}
and the Ricci tensor reads:
\begin{equation}
R_{i\bar j}=R^{\bar k}_{\ i\bar k\bar j}=-\partial_{ z_i}\Gamma^{\bar k}_{{\bar j}\bar{k}}\
=-(m+1)[\frac{\delta_{ij}}{1-|Z|^2}+\frac{\bar z_i{z}_j
 }{(1-|Z|^2)^2}]=-(m+1)g_{i\bar{j}},
\end{equation}
and the scalar curvature reads:
\begin{equation}\label{cur}
R=g^{i\bar{j} }R_{i\bar{j}}=-(m+1).
\end{equation}

We could easily verify that the metric $g_{i\bar j}$ is a
solution to the Einstein's equation in the vacuum:
\begin{equation}\label{Eevb}
R_{i\bar j}-\frac{1}{2}g_{i\bar j} R+\Lambda g_ {i\bar j}=0
\end{equation}
and the cosmological constant reads
\begin{equation}
\Lambda=\frac{m+1}{2}.
\end{equation}

\section{The discrete series of representations of $SU(m,1)$}
The Group $SU(m,1)$ possesses the principal, discrete, and
supplementary series of unitary irreducible representations, see e.g., \cite{knapp, ottoson}. The discrete series
is given by: 
\begin{equation}\label{dis}
\hat T(g) f(Z)=[\det(cZ+d)]^{-N}f(Z'),  \ N=m+1, m+2,\cdots
\end{equation}
where \begin{equation}
Z'=(aZ+b)(cZ+d)^{-1},
\end{equation} and $f(Z)$ is an arbitrary vector in the holomorphic Hilbert space $\mathcal {L}^2_N(D)$ with the measure:
\begin{equation}\label{measureu1}
d\mu_N(Z,\bar Z)= c_N [1-|Z|^2]^{N-(m+1)}|dZ|.
\end{equation}
Here $|dZ|$ is the Lebesgue measure for the complex space ${\bf C}^m$.
We choose the normalization constant
as $c_N=\pi^{-2}(N-2)(N-1)$
so that
\begin{equation}
 \int d\mu_N(Z,\bar Z)=1.
\end{equation}

In the following we consider only the case of $m=2$, as in this case the Bergman
domain $D=SU(2,1)/U(2)$ has real dimension 4 and might be more
interesting for real physical system. Remark that our method in the
follows could be easily generalized to arbitrary $m$.

Below we construct an oscillator realization of most degenerate
discrete series representations depending on one the number $N$ (see also \cite{GPW}).

We introduce a $3\times 1$ matrix $\hat Z=(\hat{z}_{a})$, $a=1, 2, 3$ of harmonic oscillators
acting in Fock space and satisfying commutation relations:
\begin{equation}\label{commu1}
 [\hat{z}_{a},\hat{z}^\dagger_{b}]=\Gamma_{ab},\  a, b=1, 2 , 3
\end{equation}
\begin{equation}
[\hat{z}_a,\hat z_b]=[\hat{z}^\dagger_a,\hat z^\dagger_b]=0,
\end{equation}
where $\Gamma$ is a $3\times 3$ matrix defined in (\ref{def}). 

It can be easily seen that for all  $g\,\in\,SU(2,1)$ these commutation
relations are invariant under transformations:
\be\label{grosilator} \hat{Z}\ \mapsto\ g\,\hat{Z},\ \ \ \hat{Z}^\dagger\
\mapsto\ \hat{Z}^\dagger\,g^\dagger .\ee
Since, $\Gamma\,=\,\mbox{diag}(+1,+1,-1)$ the upper two rows in
$\hat{Z}$ corresponds to annihilation operators whereas the lower
one to creation  operators:
\be\label{commu2}  
\hat{Z}\ =\ \left(\begin{array}{c}\hat{a}\\
\hat{b}^\dagger\end{array}\right):\ \
[\hat{a}_{\alpha},\hat{a}^\dagger_{\beta}]\ =\
\delta_{\alpha\beta}, \alpha,\beta=\,1,2.  \ \
[\hat{b},\hat{b}^{\dagger}]\ =\ 1,
 \ee
and all other commutation relations among oscillator operators
vanish. Here $\hat a$ represents a column of two oscillators $\hat a_1$ and $\hat a_2$.

The Fock space  ${\cal F}$ in question is generated from a
normalized vacuum state $|0\rangle$, satisfying
$\hat{a}_{\alpha}\,|0\rangle\,=\,\hat{b}\,|0\rangle\,=\,0$,
by repeated actions of creation operators:
\be\label{2.3u1} |m_{\alpha},\,n\rangle\ =\
\prod_{\alpha}\,\frac{(\hat{a}^\dagger_{\alpha})^{m_{\alpha}}\,(\hat{b}^\dagger)^
n}{\sqrt{m_{\alpha}!\,n!}}\,|0\rangle\,.\ee
\vskip0.5cm
We shall use the terminology that the state
$|m_{\alpha},\,n\rangle$ contains $M=\sum
m_{\alpha}$ particles  $a$ and $n$
particles  $b$.

The Lie algebra $su(2,1)$ acting in Fock space can be realized in
terms of oscillators. Consider a matrix basis of $su(2,1)$ Lie algebra:
$X=\,(X^{A}_{ab})$, where $X^A$, $A=1, \cdots 8$, is a $3\times3$ matrix, $a,b=1,2,3$, we assign the operator:
\be\label{2.3au1} 
\hat{X}^A\,=\,-\tr\hat{Z}^\dagger\Gamma
X^A\hat{Z}\ =:
-\sum_{ab}\hat{z}^\dagger_{a}\,\Gamma_{ab}\,X^A_{ab}\,\hat{z}_{b}\,,\ee
with $\hat{Z}^\dagger$ and $\hat{Z}$ defined in (\ref{commu2}).
Taking into account formula \eqref{lie} we can easily find the anti-hermicity of the operator $\hat X$:
$$ \hat{X^A}^\dagger\ =\ -\tr(\hat{Z}^\dagger
X^\dagger\Gamma \hat{Z})\ =\ +\,\tr(\hat{Z}^\dagger\Gamma
X\hat{Z})\ =\ -\,\hat{X}^A. $$

Using commutation relations for annihilation and creation operators the
commutator of operators $\hat{X}^A\,=-\,\tr\ \hat{Z}^\dagger\Gamma
X^A\hat{Z}$ and $\hat{Y}\,=-\tr\,\hat{Z}^\dagger\Gamma
Y^B\hat{Z}$ can be easily calculated:
\be\label{2.4u1} [\hat{X}^A,\hat{Y}^B]\ =\
\ [\tr\ \hat{Z}^\dagger\Gamma
X^A\hat{Z},\tr\ \hat{Z}^\dagger\Gamma Y^B\hat{Z}]\ =\
-\tr\ \hat{ Z}^\dagger\Gamma[X^A,\ Y^B]\hat{Z}\,.\ee
So we find that the operators
$\hat{X}^{A}$ satisfy in the Fock space the $su(2,1)$ commutation
relations.  The assignment
 \be\label{2.4au1}
g\,=\,e^{\xi_{A}X^{A}}\,\in\,SU(2,1) \Rightarrow\
\hat{T}(g)\,=\,e^{\xi_{A}\hat{X}^{A}} \ee
then defines a unitary  representation of the group $SU(2,1)$ in the Fock space. The explicit expressions of the operators $X^A$ are given in the appendix.

The adjoint action of $\hat{T}(g)$ on the operators $\hat Z$ reproduces (\ref{grosilator}). In terms of $a$ and $b$-oscillators in block-matrix notation this can be rewritten as
\be\label{ajoint}  g=\left(\begin{array}{cc}a&b\\c&d\end{array} \right):\ \begin{array}{cc} \hat{T}(g)\,\hat{a}\, \hat{T}^\dagger(g)\,=\,a\,\hat{a}+b\,\hat{b}^\dagger, & \hat{T}(g)\,\hat{a}^\dagger\hat{T}^\dagger(g)\,=\,\hat{a}^\dagger\,a^\dagger+\hat{b}\,b^\dagger,\\ \hat{T}(g)\,\hat{b}^\dagger\,\hat{T}^\dagger(g)\,=\,d\,\hat{b}^\dagger+c\,\hat{a}, & \hat{T}(g)\,\hat{b}\,\hat{T}^\dagger(g)\,=\,\hat{b}\,d^*+\hat{a}^\dagger\,c^\dagger,\end{array}\ee
%
in Fock space.
Since any $g\,\in\,SU(2,1)$ possesses the Cartan decomposition (\ref{cartan}) we shall discuss separately the rotations and the boosts (the action of the noncompact elements of g) given in (10). For rotations we obtain a mixing of annihilation and creation operators of a same type:
\be \label{subgroup}k=\left(\begin{array}{cc} k'&0\\0&k^{\prime\prime}\end{array}\right):\
\begin{array}{cc} \hat{T}(k)\,\hat{a}\,\hat{T}^\dagger(k)\,=\,k'\,\hat{a}, & \hat{T}(k)\,\hat{a}^\dagger\hat{T}^\dagger(k)\,=\,\hat{a}^\dagger\,k^{\prime\dagger},\\ \hat{T}(k)\,\hat{b}^\dagger\,\hat{T}^\dagger(k)\,=\,
k^{\prime\prime}\,\hat{b}^\dagger, &\hat{T}(k)\,\hat{b}\,\hat{T}^\dagger(k)\,=\,\hat{b}\,k^{\prime\prime\dagger}.\end{array} \ee
where $k'\in U(2)$, $k''\in U(1)$ and $\det(k' k'')=1$.

For a choice of the boost (for example, the operator generated by $X_7$ given in the appendix):
\begin{equation}
 \delta=\begin{pmatrix}1&0&0\\0&\cosh t&\sinh t\\0&\sinh t&\cosh t\end{pmatrix},
\end{equation}
 we have the Bogolyubov transformations for
for $\hat a_2$ and $\hat b$:
\begin{eqnarray}\label{bogou1}
 \hat{T}(\delta)\,\hat{a}_1\,\hat{T}^\dagger(\delta)&=&\hat{a}_1,\ \ \ \
 \hat{T}(\delta)\,\hat{a}_2\hat{T}^\dagger(\delta )=\cosh t
 \hat{a}_2+\sinh t \hat{b}^\dagger,\nonumber\\
\hat{T}(\delta )\,\hat{b}^\dagger\, \hat{T}^\dagger(\delta )&=&\sinh
t\hat{a}_2+\cosh t\hat{b}^\dagger.
\end{eqnarray}
Remark that $\hat{a}_1$ doesn't change under this Bogolyubov transformation.


Using the explicit form of matrices $X^{A}$ (see the appendix), following from (\ref{ajoint}), the action of generators
can be described in terms creation and annihilation of $a$- and
$b$-particles:

(i) The action of {\it rotation} generators results in a replacement of some $a_{1,2}$-particle by an other $a_{1,2}$-particle and by replacement of $b$-particle by other $b$-particle.

(ii)  The action of {\it boost} generators create or destruct of certain pair of $(a_\alpha b)$ of particles,
where $\alpha=1, 2$.

In this context it is useful to consider lowering and rising
operators  that annihilate and create $a_\alpha b$ pairs: \be\label{2.5au1}
T_-\ =\ \hat{a}_{\alpha}\,\hat{b},\ \ \ T_+={(T_- )}^\dagger =\
\hat{b}^\dagger \hat{a}_\alpha^\dagger,\ee
such that any boost given by (\ref{2.3au1}) of a complex combination of $\hat X^A$, $A=4,5,6,7$,
can be uniquely expressed as complex combinations of operators $T_-$ and $T_+$.


It follows from (\ref{2.5au1}) that the operator
\be\label{2.6u1}   \hat{N}\equiv\ \hat{N}_{\hat{b}}-\hat{N}_{\hat{a}}\
=\ \,\hat{b}^\dagger\hat{b}
-\hat{a}^\dagger_{1}\hat{a}_{1}-\hat{a}^\dagger_{2}\hat{a}_{2} =-[
\hat{Z}^\dagger\Gamma\hat{Z}+1] \ee
commutes with all generators $\hat{X}^{A}$. \vskip0.5cm

Below, we shall restrict ourselves to most degenerate discrete series representations which are specified by the eigenvalue of the operator $ \hat{N}$  in the representation subspace.  We start to construct the representation space ${\cal F}_N$ from a distinguished normalized state containing lowest number of particles:
\be\label{2.7u1} |z_0\rangle = \frac{(\hat{b}^\dagger)^{N}
}{\sqrt{(N)!}}|0> =
\frac{1}{\sqrt{N}}\,|0,0;N\rangle\,.\ee
Here $N$ is a natural number that specifies the representation: $ \hat{N}\,|z_0\rangle\,=\,N\,|z_0\rangle$. All other states in the representation space are obtained by the action of rising operators given in (\ref{2.5au1}): such states contain besides $N$ $b$-particles a finite number of $a b$ pairs.\vskip0.5cm

The stability group of the state $|z_0\rangle$ is maximal compact subgroup $K\,=\,S(U(2)\times U(1))$. The group action for $k\,=\,\mbox{diag}(k',k^{\prime\prime})\in K$ reduces just to the phase transformation  (see (\ref{grosilator}) or  (\ref{subgroup})):
\be\label{2.7au1} \hat{b}^\dagger\ \mapsto\ \hat{b}^\dagger\,k^{\prime\prime}\ \ \Rightarrow\ \  |z_0\rangle\ \mapsto\ e^{iN\alpha } \, |z_0\rangle\ee
where $k^{\prime\prime}=e^{i\alpha(k)}$ is the $U(1)$ part of $K$.

Define the phase function as $\omega_0(g) = \langle
z_0|\,\hat{T}(g)\,|z_0\rangle$. Using the decomposition $g = k^{\dagger}\,\delta\,q$  and the
action of rotations (\ref{2.7au1}) we obtain:
\begin{eqnarray}
 \omega_0(g)&=& \langle z_0|\,\hat{T}(g)\,|z_0\rangle\ =\ \langle z_0|\,\hat{T}(k)^\dagger\,\hat{T}(\delta)\,\hat{T}(q)\,|z_0\rangle\nonumber\\
&=&e^{iN(\alpha(q)-\alpha(k))} \langle z_0|\,\hat{T}(\delta)\,|z_0\rangle, 
\end{eqnarray}
where $\alpha(k)$ and $\alpha(q)$ are the phase factors associated to the compact subgroups $k$ and $q$.

Since the compact part only results in a phase factor, it is enough to calculate the mean value for a noncompact
element:
\begin{eqnarray}\label{boost}
&&\delta=\begin{pmatrix}1&0&0\\0&\cosh t&\sinh t\\0&\sinh t&\cosh
t\end{pmatrix}=t_+\  t_0\  t_-\\&=&\begin{pmatrix}1&0&0\\0&1&\tanh t\\0&0&1\end{pmatrix}\begin{pmatrix}1&0&0\\0&\cosh^{-1} t& 0\\0&0&\cosh t\end{pmatrix}
\begin{pmatrix}1&0&0\\0&1&0\\0&\tanh t&1\end{pmatrix}\nonumber
\end{eqnarray}

%

In the representation in question, the matrices $t_+$ and $t_-$ are the
exponents of the matrices
\be
 X_+=\begin{pmatrix}0&0&0\\0&0&\tanh t\\0&0&0\end{pmatrix}, \  X_-=\begin{pmatrix}0&0&0\\0&0&0\\0&\tanh t&0\end{pmatrix},
\ee
respectively.
Therefore we have
\begin{equation}
  \hat{T}(t_+)\,=\, e^{\hat{a}_2^\dagger\,\tanh t\,\hat{b}^\dagger},\ \ \hat{T}(t_-)\,=\, e^{b\,\tanh t\,a_2}.
\end{equation}

Since $ \hat{T}(t_-)$  contains  $\hat{a_2}$ its action does not
affect  $|z_0\rangle$, and similarly $\hat{T}(t_+)$ containing
$\hat{a_2}^\dagger$ does not affect $\langle z_0|$. The only
non-trivial action comes from
\be\label{2.9u1} \hat{T}(t_0)\,=\,
e^{-b\hat{b}^\dagger \ln \cosh t+
\hat{a_2}\hat{a_2}^\dagger\ln\cosh t
}=\frac{1}{\cosh t}e^{-\hat{b}^\dagger b \ln \cosh t+
\hat{a_2}\hat{a_2}^\dagger\ln\cosh t
}.\ee
Since the actions of the operators $\hat a_2$ and $\hat{a_2}^\dagger$ on the state $|z_0\rangle$
are trivial, we have
\begin{eqnarray}\label{2.9a}
&&\hat{T}(t_0)|z_0\rangle=\frac{1}{\cosh t}e^{-\hat{b}^\dagger b \ln \cosh t}|z_0\rangle\\
&=&\frac{1}{\cosh t}\sum_{n=0}^N (-1)^n\frac{(\ln\frac{1}{\cosh t})^n}{ n!}(b^\dagger)^n b^n\frac{(b^\dagger)^N}{\sqrt{(N)!}}|0\rangle\nonumber\\
&=&\frac{1}{\cosh t}\sum_{n=0}^N(-1)^n C^n_N (\ln\frac{1}{\cosh t})^n
\frac{(b^\dagger)^N}{\sqrt{(N)!}}|0\rangle=\frac{1}{\cosh
t}(1+\ln\cosh t)^N|z_0\rangle\nonumber
\end{eqnarray}

From equations (\ref{2.7au1}) and  (\ref{2.9u1}) we obtain:
\be\label{2.10} \omega_0(g) = \langle z_0|\,\hat{T}(g)\,|z_0\rangle
= \frac{1}{\cosh t}[(1+\ln\cosh t)e^{i(\alpha(q)-\alpha(k))}]^N \ee.

%

\section{The Coherent State Quantization and the Star Product}

 We briefly describe the construction of coherent states \`a la Perelomov \cite{Perelomov}.

Let $T_g$ be an unitary irreducible representation of an arbitrary Lie group $G$ in a Hilbert space $\mathcal{H}$,
$|z_0\rangle\in\mathcal{H}$ is a normalized state in the Garding space of $T_g$.
Let $K$ be the stability group of the $|z_0\rangle$ such that 
$T_k|z_0\rangle=e^{i\alpha(k)}|z_0\rangle,\ \ {\rm for}\  k\in K$. Then for each boost $g_z$ and point $z=g_z z_0\in D=G/K$ we could assign a coherent states :
$ |z\rangle=\psi_z= T(g_z)|z_0\rangle$.
Define also the functions $\omega_0(g)=<z_0|T(g)|z_0>$ and $\omega (g,z)=<z|T_g|z>=\omega_0 (g_z^{-1}g g_z) $. As $|z_0\rangle$ is in the Garding space, $\omega(g)$ is a smooth function in $g$.

To any boost $g_z$ defined in (\ref{boost}), we define coherent coherent state (see,
\cite{Perelomov}) as:
\be\label{2.12} |z\rangle\ =\  \hat{T}(g_z)\,|z_0\rangle\ =\
\hat{T}(k\,\delta\,k^\dagger)\,|z_0\rangle,\ \   z = z(k,\delta).
\ee
Let us consider operators in the representation space of the form
\be\label{2.13} \hat{F}\ =\ \int_G dg\,\tilde{F}(g)\, \hat{T}(g),\ee
where $\tilde{F}(g)$ is a distribution on a group $G$ with compact
support. To any such operator we assign
function on $G/K$ by the prescription
\be\label{2.14} F(Z)\ =\ \langle z|\,\hat{F}\,|z\rangle\ =\ \int_G
dg\,\tilde{F}(g)\,\omega(g,z),\ee
where
\be\label{2.14a} \omega(g,z)\ \equiv\ \langle
z|\,\hat{T}(g)\,|z\rangle\ =\  \omega_0(g^{-1}_z\,g\,g_z).\ee
This equation combined with (\ref{2.10}) offers an explicit form of
$\omega(g,x)$ and is well suited for calculations.
Similarly we can define the function
\be\label{2.14} G(Z)\ =\ \langle z|\,\hat{G}\,|z\rangle\ =\ \int_G
dg\,\tilde{G}(g)\,\omega(g,z),\ee

Due to the non-commutativity of $\hat T(g)$ the product of the two functions $F(Z)$ and $G(Z)$ are no longer commutative. The star-product of two functions is defined as \cite{GP}:
\be (F\star G)(Z)\label{stardef} =\ \langle z|\,\hat{F}\hat{G}\,|z\rangle\ =\ \int_{G\times G} dg_1 dg_2\,\tilde{F}(g_1)\,\tilde{G}(g_2)\, \omega(g_1g_2,z)\ee
\be\label{2.15} =\ \int_G
dg\,(\tilde{F}\circ\tilde{G})(g)\,\omega(g,z),\ee
where the symbol $\tilde{F}\circ\tilde{G}$ denotes the
convolution in the group algebra $\tilde{\cal A}_G$ of
distributions with compact support:
\be\label{2.16} (\tilde{F}\circ\tilde{G})(g)\ =\   \int_G
dh\,\tilde{F}(gh^{-1})\,\tilde{G}(h),\ee
We can easily find that the star product defined above is associative and is invariant
under the action of G.
It is easy to find that:
\bea&& \mbox{supp}\,(\tilde{F}\circ\tilde{G})\ \subset\ (\mbox{supp}\,\tilde{F})\,(\mbox{supp}\,\tilde{G})\nonumber\\
&\equiv&\ \{ g\,=\,g_1 g_2\,|\, g_1\in\mbox{supp}\,\tilde{F},\,g_2\in \mbox{supp}\,\tilde{G}\}. \eea
consequently, for a non-compact group there are two classes of group
algebras:

(i) The first one is generated by distributions with a general
compact support and the corresponding group algebra is simply the
full algebra $\tilde{\cal A}_G$ defined in (\ref{2.15}).

(ii) The second one is formed by  distributions $\tilde{F}$ with
$\mbox{supp}\,\tilde{F}$ subset of a subgroup $H\subset K$, form a
sub-algebra $\tilde{\cal A}_H$ of  the group algebra $\tilde{\cal
A}_G$.

{\it Note}: We point out that as in the case of usual distributions,
the convolution product may exist even for distribution with
non-compact support provided there are satisfied specific
restriction at infinity.  \vskip0.5cm

In the second class there are two interesting cases:

(a) $\tilde{\cal A}_{\{e\}}$ corresponding to the trivial subgroup
$K = \{e\}$ in $G = SU(2,1)$.

(b) $\tilde{\cal A}_K$ corresponding to the maximal compact subgroup
$K$ in $G$, which corresponds to the complex ball.

{\it Note}: It is well-known that $\tilde{\cal A}_{\{e\}}$ is
isomorphic to the enveloping algebra ${\cal U}(su(2,1))$ (see e.g.,
\cite{Kirillov}). \vskip0.5cm



Any distribution  $\tilde{F}$ can be given as a linear combination
of finite derivatives of the group $\delta$-function, i.e., as a
linear combination of distributions
\be\label{2.17}  \tilde{F}_{A_1\dots A_n}(g)\ =\ ({\hat
X}_{A_1}\,\dots\, {\hat
X}_{A_n}\delta)(g),\ee
where ${\hat
X}_{A_i}$ with $A_i=1, \cdots, 8$ are the left-invariant vector fields on group $G$
representing the Lie algebra generators $X_{A_i}$ whose explicit form is given in the appendix. Here we will not distinguish the upper indices and lower indices of 
the Lie algebra elements and will identify
$X_A$ with $X^A$.
Inserting this into
(\ref{2.14}) we obtain the corresponding function from ${ \cal
A }^\star_{\{e\}}$
\be\label{2.18}  F_{A_1\dots A_n}(Z)\ =\ (-1)^n ({\hat
X}_{A_n}\,\dots\, {\hat X}_{A_1}\omega)(g,z)|_{g=e}.\ee
Here we used the fact that the operators ${\hat
X}_{A}$ are
anti-hermitian differential operators with respect to the group
measure $dg$. From (\ref{2.14}) it follows directly that
$$ (F_{A_1\dots A_n}\star F_{B_1\dots B_m})(Z) $$
\be\label{2.19}   =\ (-1)^{n+m}({\hat X}_{A_n}\,\dots\, {\hat
X}_{A_1}\,{\hat
X}_{B_m}\,\dots\, {\hat
X}_{B_1}\omega)(g,Z)|_{g=e}.\ee
Equations (\ref{2.18}) and (\ref{2.19}) describe explicitly the
homomorphism ${\cal U}(su(2,1))\to{\cal A}^\star_{\{e\}}$.
\vskip0.5cm

Using exponential parametrization of the group element $g =
e^{\xi^{A}X_{A}}$ formula for the symmetrized function
(\ref{2.18}) takes simple form:
$$ F_{\{A_1\dots A_n\}}(Z)\ =\ (-1)^n (\partial_{\xi_{A_1}}\,\dots\, \partial_{\xi_{A_n}}\omega)( e^{\xi^{A}\hat X_{A}}Z)|_{\xi=0} $$
\be\label{2.20} =\ (-1)^n\,\langle z|\,\hat{X}_{\{A_1}\,\dots\,
\hat{X}_{A_n\}}\\|z\rangle, \ee
where $\{\,\dots\,\}$ means symmetrization of double indexes and
$\xi=0$ means the evaluation at $\xi_{A}=0$ for $A_n=1,\dots,8$.
Symmetrized functions form a basis of the algebra in question and symmetrized
elements from the center of algebra correspond to Casimir operators.
In the series of representation in question all Casimir operators
are given in terms of a single operator $\hat{N}$ given in
(\ref{2.6}) which is represented by a constant function $N(x) =
\langle x|\hat{N}|x\rangle = N$. 

For case b) we simply take the indices of $\xi_{A_i}$ as $A_i=5, 6, 7 , 8$, when we take the basis of the Lie algebra as that in appendix B. \vskip0.5cm

The boosts of $g_z$ can be chosen as the form:
\begin{equation}\label{2.11} g_z =  k\,\delta\,k^\dagger\ =\begin{pmatrix}
C'&S'\\S'^\dagger&C''\\ \end{pmatrix}\in G/K,
\end{equation}
where $C'=k'\begin{pmatrix}1&0\\0&\cosh t\\  \end{pmatrix}k'^\dagger
$, \ $S'=\begin{pmatrix} k'_{12}\sinh t\\k''_{22}\sinh t\\  \end{pmatrix}k''^\star
$,\  $C''=\cosh t$. $k'$ and $k''$ are defined by formula \eqref{subgroup}.

Let us calculate the function $\omega(g,z) = \omega_0(g_z^{-1}\,g\,g_z)$
explicitly. Taking $g$ and $g_z$ as in (\ref{cartan}) and (\ref{2.11})
we have to calculate the product of three matrices:
$$ g_z^{-1}\,g\,g_z\ =\
\left(\begin{array}{cc}C' & -{S'}\\ -{S'}^\dagger &
C^{\prime\prime}\end{array}\right)
\left(\begin{array}{cc}a&b\\c&d\end{array}\right)
\left(\begin{array}{cc}C' &S\\ S^\dagger &
C^{\prime\prime}\end{array}\right).$$
Using equation (\ref{2.10}) for $\omega_0(g)$ we obtain
\begin{eqnarray}\omega(g,z)&=& \mbox{det}[C'a C'\,+\,k''C' b\
(\bar{k'}_{12},\ \bar{k'}_{22})\sinh t-\,\sinh t
\bar{k''}\begin{pmatrix}k'_{12}\\k'_{22}\\\end{pmatrix}\ c\
C'\nonumber\\&-&d\sinh^2 t
\begin{pmatrix}k'_{12}\\k'_{22}\\\end{pmatrix}(\bar{k'}_{12},\  \bar{k'}_{22})]^{-N}
\end{eqnarray}
After some basic calculations we have:
\begin{eqnarray}\label{2.20} \omega(g,z)&=& (\cosh t)^{-2N}\,\mbox{det}[a\,+\,k''\tanh t b\
(\bar{k'}_{12},\  \bar{k'}_{22})-\,\tanh t
\bar{k''}\begin{pmatrix}k'_{12}\\k'_{22}\\\end{pmatrix}\  c
\nonumber\\&-&d\tanh^2 t
\begin{pmatrix}k'_{12}\\k'_{22}\\\end{pmatrix}(\bar{k'}_{12},\
\bar{k'}_{22})]^{-N}.
\end{eqnarray}
\vskip0.5cm

We define the function $\xi_A$ as the expectation value of the operator $\hat{X}_A$ between the coherent states as;
\be\label{nccoord} \xi_{A}(z) = \frac{1}{N}\,\langle
z|\hat{X}_{A}|z\rangle = \frac{1}{N}\,\langle z_0|\hat{T}^\dagger
(g_z)\,\hat{X}_{A}\hat{T}(g_z)|z_0\rangle, \ee
for $A=\,1,\,\dots,\,8$. Taking into account (\ref{ajoint}) we
see that $\xi_{A}(z) = D^{B}_{A}(g_z)$, where $Ad^*_g =
(D^{B}_{A}(g))$ is the matrix corresponding to the group action in
co-adjoint algebra. Therefore it is sufficient to evaluate the
coordinates at $z_0$: $\xi_{A}(z_0) = \frac{1}{N}\,\langle
z_0|\hat{X}_{A}|z_0\rangle$.
\vskip0.5cm

The star product between these functions are defined as
\be\label{starcoo}
\xi_A \star \xi_B=\frac{1}{N^2}\,\langle
z|\hat{X}_{A}\hat{X}_{B}|x\rangle,
\ee
or more explicitly:
\begin{equation}
 (\xi_{A}\star\xi_{B})(z)\ =\ \frac{1}{2N^2}\,\langle
z|\{\hat{X}_{A},\hat{X}_{B}\}|z\rangle\ +\ \frac{1}{2N^2}\,
\langle z|[\hat{X}_{A},\hat{X}_{B}]|z\rangle ,
\end{equation}

where $\{\,\dots\,\}$ denotes anti-commutator and $[\,\dots\,]$ is
commutator. Therefore, the second term is
\begin{equation} \frac{1}{2N^2}\,\langle z|[\hat{X}_{A},\hat{X}_{B}]|z
\rangle\ =\ \frac{1}{2N}\,f^{C}_{A,B}\,\xi_{C}(z),
\end{equation}
where we used the definition of $\hat{X}_{A}$ and the short-hand
notation for the commutator: $[X_{A},X_{B}] =
f^{C}_{A,B} X_{C}$. The first term is proportional to the
symmetrized function $F_{\{A,B\}}$ and we can use (\ref{2.20}):
\begin{equation}
  \frac{1}{2N^2}\,\langle z|\{\hat{X}_{A},\hat{X}_{B}\}
|z\rangle\ =\ (1 + A_N)\,\xi_{A}(x)\,\xi_{B}(z)\,+\,B_N\,
\delta_{AB},
\end{equation}
where we have a usual point-wise product of functions in the first
term.

The coefficients $A_N$ and $B_N$ depend on the Bernoulli numbers coming from the Baker-Campbell-Hausdorff
formula and are of order $1/N$.

So that we have:
\be\label{star2} (\xi_{A}\star\xi_{B})(z)\ =\ (1+ A_N)\,
\xi_{A}(z)\,\xi_{B}(z) \,+\,\frac{1}{2N}\,f^{C}_{A,B}
\,\xi_{C}(z)\,+\,B_N\, \delta_{A,B}. \ee
The Harish-Chandra imbedding theorem states that we could always
imbed the commutative maximal Hermitian symmetric space into the noncompact part of
the Cartan subalgebra. So it is natural to consider the (noncommutative) functions $\xi_{A}$ with $A=4,5,6,7$
 as the coordinates of the noncommutative complex ball $\hat D$.

We see that the parameter of the non-commutativity is $\lambda_N =
1/N$. For $N\,\to\,\infty$ we recover the commutative product.

\section{A scalar field model in $\hat D$}
In this section we consider a quantum theory of scalar fields on the
noncommutative complex ball. 
The Lagrangian is defined as
\begin{equation}\label{model}
 S[\Phi]=\int d\mu(Z,\bar Z)\{-\frac{1}{2}\Phi\star\Delta_N \Phi (Z,\bar Z)-\frac{1}{2}[\mu^2+\xi R] \Phi^2(Z,\bar Z)_\star-g\Phi^4_\star\}
\end{equation}
where $d\mu(Z,\bar Z)$ is the invariant measure of $D$ given by (\ref{measureu1}), $\Delta_N$ is the invariant Laplacian operator labeled by a positive integer $N$ (see Appendix C for more details);
The complex scalar fields $\Phi$  {\footnote{In order that $\Phi$ be a $scalar$ hence the full action to be $SU(m,1)$ invariant, the function space of $\Phi$ should be the Moebius-invariant subspace of the full representation space of $SU(m,1)$. A space Y of functions with domain D is said to be Moebius invariant if $T_g\Phi\in Y$ for all $\Phi\in Y$ and $T_g\in Aut(D)$. The explicit construction of the Mobius invariant functions could be found in \cite{rudin}.}} are unitary irreducible representations of
$SU(m,1)$ in the Hilbert space, analogous to the Minkowskian quantum field theories where the fields are unitary irreducible representations of the Poincar\'e group. They can be considered as polynomial
functions of the noncommutative coordinates $\xi_A$ (see, \eqref{nccoord}); $\mu$ is the mass of $\Phi$,
$R=-(m+1)$ is the curvature scalar, see (\ref{cur}), and $\xi$ is a non-vanishing
numerical factor.

Remark that the most general invariant Laplacian operator is written as $\Delta_\nu$ where
$\nu>-1$ is a real number. The fact that we take $\nu=N$ is exactly due to that we consider only the discrete series of representation (see; e.g. \cite{PZh}) and N is defined in \eqref{dis}.  

Let $\phi(N, l, Z)$ be the
eigenfunction of $\Delta_N$ labeled by non-negative integers $l$, so we have:
\begin{equation}
\Phi=\sum_{l=0}^{[\frac{N-2}{2}]} C_{N,l}\phi(N, l,Z),
\end{equation}
where $l$ label
the discrete series and $C_{N,l}$ are complex numbers.

Let $F[\Phi]$ be the polynomial function of the field $\Phi$, then
the quantum expectation value of 
$F[\Phi]$ is defined as the functional integral over fields
$\Phi$:
\begin{equation}
 \langle F[\Phi]\rangle\ =\ \frac{\int D[\Phi]\,e^{-S[\Phi]}\,
 F[\Phi]}{\int D[\Phi]\,e^{-S[\Phi]}},
\end{equation}
where
\begin{equation}
\int D[\Phi]=\int_D \prod_{Z,\bar Z\in D} d\Phi(Z,\bar Z).
\end{equation}

The spectrum of  $\Delta_N$ is given in Appendix C and we can easily find that the noninteracting propagator reads:
\begin{equation}
<\Phi, \Phi>= \frac{1}{\mu^2-(m+1)\xi+\frac{1}{4}((N-2)^2+\lambda^2)},
\end{equation}
where in general $\lambda$ can be an arbitrary complex valued function. 
For the discrete series we have (see \cite{PZh}) $ \lambda=i(N-2-2l)$,  $l=0, 1, \cdots
,\big[\frac{N-2}{2}\big]$. For the $m=2$ case the propagator reads:
\begin{equation}
 \frac{1}{l(N-2)-l^2+\mu^2-3\xi}.
\end{equation}

Now we consider the interaction vertex $g\int d\mu(Z,\bar Z)\Phi^4_\star$. From the discussion of the previous section \eqref{star2} we find that the stat product could be written as the commutative product plus the noncommutative corrections of order $1/N$, for the lowest order approximation. So we can consider the correlation
functions for the noncommutative scalar model as the commutative ones plus the noncommutative $1/N$ corrections. 

The treatment of the quantum field theory model in this paper is very sketchy. The explicit expression of the vertex functions and a systematic study of the quantizations and renormalizations of this model deserve another paper.
Even the quantization of the scalar field theory on the commutative complex ball is not worked out in the literature. 
We shall study all these in future publications.

\section{Appendix A, the Lie algebra $su(2,1)$}
The Lie algebra $\textbf{g}=su(2,1)$ is formed by $3\times3$ complex
matrices $X$ satisfying
\be\label{1.1b0} X^\dagger\,\Gamma\ +\ \Gamma\,X\ =\ 0.\ee

It has $8$ independent generators that form an orthogonal complete
basis. We could choose them as:
\begin{equation}
 X_1=\begin{pmatrix}
0 & i & 0 \\
i & 0 & 0\\
0 & 0 & 0\\
\end{pmatrix},
 X_2=\begin{pmatrix}
0 & 1 & 0 \\
-1 & 0 & 0\\
0 & 0 & 0\\\end{pmatrix},
 X_3=\begin{pmatrix}
i & 0 & 0 \\
0 & -i & 0\\
0 & 0 & 0\\\end{pmatrix},\nonumber
\end{equation}
\begin{equation}
 X_4=\begin{pmatrix}
0 & 0 & i \\
0 & 0 & 0\\
-i & 0 & 0\\\end{pmatrix},
 X_5=\begin{pmatrix}
0 & 0 & 1 \\
0 & 0 & 0\\
1 & 0 & 0\\\end{pmatrix},
 X_6=\begin{pmatrix}
0 & 0 & 0 \\
0 & 0 & i\\
0 & -i & 0\\\end{pmatrix},\nonumber
\end{equation}
\begin{equation}
 X_7=\begin{pmatrix}
0 & 0 & 0 \\
0 & 0 & 1\\
0 & 1 & 0\\\end{pmatrix},
 X_8=\frac{1}{\sqrt{3}}\begin{pmatrix}
i & 0 & 0 \\
0 & i & 0\\
0& 0 & -2i\\\end{pmatrix},
\end{equation}

Under this basis the infinitesimal generators could be written as:
\begin{eqnarray} \label{basis}
\hat X_1&=&i(z_1\frac{\partial}{\partial z_2}+z_2\frac{\partial}{\partial z_1}),\  \hat X_2=-z_1\frac{\partial}{\partial z_2}+z_2\frac{\partial}{\partial z_1},\  \hat X_3=i(z_1\frac{\partial}{\partial z_1}-z_2\frac{\partial}{\partial z_2}),\nonumber\\
\hat X_4&=&i\frac{\partial}{\partial z_1}+iz_1^2 \frac{\partial}{\partial z_1}+iz_1 z_2\frac{\partial}{\partial z_2},\ \hat X_5=\frac{\partial}{\partial z_1}-z_1^2 \frac{\partial}{\partial z_1}-z_1 z_2\frac{\partial}{\partial z_2}, \nonumber\\
\hat X_6&=&i\frac{\partial}{\partial z_2}+iz_1 z_2
\frac{\partial}{\partial z_1}+i z_2^2\frac{\partial}{\partial z_2},\
\hat X_7=\frac{\partial}{\partial z_2}-z_1 z_2
\frac{\partial}{\partial z_1}- z_2^2\frac{\partial}{\partial z_2},
\nonumber\\ \hat
X_8&=&-\frac{i}{\sqrt{3}}(z_1\frac{\partial}{\partial
z_1}+z_2\frac{\partial}{\partial z_2})+\frac{2n}{\sqrt{3}}i.
\end{eqnarray}
Where $Z=(z_1, z_2)^\dagger$ is the coordinate of the coset space
$D$ and $n$ is a number to characterize the representation. More
explicitly, $n=a+ib$, $a, b\in R$ for the principal series;
$n\in N$ a natural number for the discrete series and $n\in(-1/2, 0)$ for the
supplementary series of representations. \vskip1cm

\section{Appendix B, the harmonic oscillator representation for $SU(m,n)$}\label{mn}

In this section we realize the maximal degenerate
discrete series of representations for the $SU(m,n)$ group in terms of harmonic oscillators.

We introduce a $m\times n$ matrix $\hat{Z} = (\hat{z}_{a\alpha })$,
$a\,=\,1,\,\dots,\,m$, $\alpha \,=\,1,\,\dots,\,n$, of bosonic oscillators that act in Fock space and satisfy the following commutation relations:
\begin{equation}\label{2.1}
 [\hat{z}_{a\alpha },\hat{z}^\dagger_{b\beta }]\,=\,\Gamma_{ab}\,\delta_{\alpha \beta},\ \ \  a, b=1, 2,
\end{equation}
\begin{equation}
[\hat{z}_{a,\alpha}, \hat z_{b\beta}
]\,=\,[\hat{z}^\dagger_{a,\alpha},\hat z^\dagger_{b\beta}]\,=0\,,
\end{equation}
where $\Gamma$ is a $(m+n)\times (m+n)$ matrix defined in
(\ref{unit}). It
can be easily seen that these commutation relations are invariant under transformations:
\be\label{2.1a} \hat{Z}\ \mapsto\ g\,\hat{Z},\ \ \hat{Z}^\dagger\
\mapsto\ \hat{Z}^\dagger\,g^\dagger,\ \ g \in SU(m,n).\ee
%
%
 From (\ref{2.1}) it follows that the upper $m\times n$ block in
 $\hat{Z}$ corresponds to annihilation operators whereas the lower
 $m\times m$ block is formed from creation  operators.
So we rewrite the matrix $\hat Z$ as
\be\label{2.2}  \hat{Z}\,=\,\left( \begin{array}{c}\hat{a}\\
\hat{b}^\dagger \end{array}\right)\ee
such that
$$ [\hat{a}_{a\alpha},\hat{a}^\dagger_{b\beta}]\,=\,\delta_{ab}
\delta_{\alpha\beta},\ \ a,b=1,\dots,m,\ \alpha,\beta=1,\dots,n,$$
\be\label{2.2a} [\hat{b}_{a\alpha},\hat{b}^{\dagger}_{b\beta}]\,=\,\delta _{ab}\,\delta_{\alpha\beta},\ \ a,b,\alpha,\beta = 1,\dots,n.\ee
All other commutation relations between annihilation (creation) operators vanish. The Fock space  ${\cal F}$ in question is generated from a
normalized vacuum state $|0\rangle$, satisfying
$\hat{a}_{\alpha}\,|0\rangle\,=\,\hat{b}\,|0\rangle\,=\,0$,
by repeated actions of creation operators:
\be\label{2.3} |m,\,n\rangle\ =\
\prod_{a\alpha,b\beta }\,\frac{(\hat{a}^\dagger_{a\alpha})^{m_{a\alpha}}\,(\hat{b}^\dagger_{b\beta})^
{n_{b\beta}}}{\sqrt{m_{a\alpha}!\,n_{b\beta}!}}\,|0\rangle\,.\ee
Here, $m = (m_{a\alpha})$ and $n = (n_{b\beta})$ are matrices of non-negative integers and
the range of indexes $a,\alpha$ and $b,\beta $ have been indicated in (\ref{2.2a}).
We shall use the terminology that the state $|m,\,n\rangle$ contains $M=\sum m_{a\alpha}$ particles  $a$ and $N=\sum n_{b\beta}$
particles  $b$.
\vskip0.5cm

The Lie algebra $su(m,n)$ acting in Fock space can be realized in
terms of oscillators. To any basis $X^{AB}\in \textbf{g}$, where $A, B=1,2 \cdots (m+n)^2-1$ we assign the {\it anti-hermitian} operator
\be\label{2.3a} \hat X^{AB}\ =\ -\mbox{tr}\,(\hat{Z}^\dagger\Gamma
X^{AB}\hat{Z})\ =\
-\sum_{a,b,c,d,\alpha}\hat{z}^\dagger_{a\alpha }\,\Gamma_{ac}\,X^{AB}_{cb}\,\hat{z}_{b\alpha }\,,\ee
with $\hat{Z}^\dagger$ and $\hat{Z}$ given in (\ref{2.1}) and
(\ref{2.2}) in terms of oscillators.

Using commutation relations for annihilation and creation operators the commutator of operators $\hat X^{AB}\,=-\tr\,\hat{Z}^\dagger\Gamma
X^{AB}\hat{Z}$ and $\hat{Y}^{CD}\,=\,-\tr\hat{Z}^\dagger\Gamma
Y^{CD}\hat{Z}$ can be easily calculated:
\be\label{2.4} [\hat{X},\hat{Y}]\ =\
[-\tr\ \hat{Z}^\dagger\Gamma
X\hat{Z},-\tr\ \hat{Z}^\dagger\Gamma Y\hat{Z}]\ =\
-\tr\ \hat{ Z}^\dagger\,\Gamma\,[X,\ Y]\,\hat{Z}\,.\ee
It can be easily seen that the
anti-hermitian operators $\hat{X}_{A}$ satisfy  the $su(m,n)$ commutation relations in Fock space. The assignment
 \be\label{2.4a}
g\,=\,e^{\xi^{A}X_{A}}\,\in\,SU(m,n)\ \Rightarrow\
\hat{T}(g)\,=\,e^{\xi^{A}\hat{X}_{A}} \ee
then defines the unitary $SU(m,n)$ representation in the Fock space.

The adjoint action of $\hat{T}(g)$ on operators reproduces the left-action (\ref{2.1a}). In terms of $a$ and $b$-oscillators in block-matrix notation this can be rewritten as
\be\label{2.4b}  g=\left(\begin{array}{cc}a&b\\c&d\end{array} \right):\ \begin{array}{cc} \hat{T}(g)\,\hat{a}\, \hat{T}^\dagger(g)\,=\,a\,\hat{a}+b\,\hat{b}^\dagger, & \hat{T}(g)\,\hat{a}^\dagger\hat{T}^\dagger(g)\,=\,\hat{a}^\dagger\,a^\dagger+\hat{b}\,b^\dagger,\\ \hat{T}(g)\,\hat{b}^\dagger\,\hat{T}^\dagger(g)\,=\,d\,\hat{b}^\dagger+c\,\hat{a}, & \hat{T}(g)\,\hat{b}\,\hat{T}^\dagger(g)\,=\,\hat{b}\,d^\dagger+\hat{a}^\dagger\,c^\dagger.\end{array}\ee
Since any $g\,\in\,SU(m,n)$ possesses Cartan decomposition (\ref{cartan}) we shall discuss separately rotations and special boosts.

For any rotation $k\in \textbf{k} = s(u(m)\oplus u(n))$ we obtain the mixing of annihilation and creation operators of the same type:
\be\label{rotation} k=\left(\begin{array}{cc} k_1&0\\0&k_2\end{array}\right):\ \begin{array}{cc} \hat{T}(k)\,\hat{a}\,\hat{T}^\dagger(k)\,=\,k_1\,\hat{a}, & \hat{T}(k)\,\hat{a}^\dagger\hat{T}^\dagger(k)\,=\,\hat{a}^\dagger\,k_2,\\ \hat{T}(k)\,\hat{b}^\dagger\,\hat{T}^\dagger(k)\,=\,
k_2\,\hat{b}^\dagger, &\hat{T}(k)\,\hat{b}\,\hat{T}^\dagger(k)\,=\,\hat{b}\,k^\dagger_2.\end{array} \ee
where $\hat{a} = (\hat{a}_{a\alpha })$, $\hat{a}^\dagger = (\hat{a}^\dagger_{a\alpha })$, $\hat{b} = (\hat{b}_{b\beta })$, $\hat{b}^\dagger = (\hat{b}^\dagger_{b\beta })$ denote matrices consisting from the corresponding oscillator operators.

However, for the special boost $\delta_\Lambda$ given in (\ref{boost}) we obtain Bogolyubov transformations. In matrix notation the Bogolyubov transformation takes the form:
\begin{eqnarray}\label{bogo}
 \hat{T}(\delta_\Lambda )\,\hat{Z}\,\hat{T}^\dagger(\delta_\Lambda )&=& \delta_\Lambda\,\hat{Z},\nonumber\\
\hat{T}(\delta_\Lambda  )\,\hat{Z}^\dagger\,\hat{T}^\dagger(\delta_\Lambda  )&=&\hat{Z}^\dagger\,\delta_\Lambda .
\end{eqnarray}
All other boosts can be obtained from the special boosts by rotations.

So as the same in the $SU(2,1)$ case, the action of {\it rotation} generators results in a
replacement of some $a$-particle by an other $a$-particle and by replacement of $b$-particle by other $b$-particle
and the action of {\it boost} generators results in a creation of
some $ab$-pair of particles or in a destruction of some other $ab$ pair.

Consider the lowering and rising
operators that annihilate and create some $a_{a\alpha}b_{b\beta}$
pair: \be\label{2.5a} T_{a\alpha,b\beta}^-\ =\
\hat{a}_{a\alpha}\,\hat{b}_{b\beta},\ \ \
T_{a\alpha,b\beta}^+={(T_{a\alpha,b\beta}^- )}^\dagger =\
\hat{b}_{b\beta}^\dagger \hat{a}_{a\alpha}^\dagger.\ee
Any boost can be uniquely expressed as complex combinations of lowering and rising operators.

It follows from (\ref{2.5a}) that the operator
\be\label{2.6}   \hat{N}\equiv\ \hat{N}_{\hat{b}}-\hat{N}_{\hat{a}}\ =\ \sum_{b\beta}\hat{b}_{b\beta}^\dagger \hat{b}_{b\beta}\,-\,  \sum_{a\alpha}\hat{a}_{a\alpha}^\dagger \hat{a}_{a\alpha}\
=\ -\mbox{tr\,}(\hat{Z}^\dagger\Gamma\hat{Z})\,-\,n^2 \ee
commutes with all generators $\hat{X}_{A}$. \vskip0.5cm

We start to construct the representation space ${\cal F}_N$ from a distinguished normalized state containing lowest number of particles:
\be\label{2.7} |z_0\rangle = c^{-1}_N\,(\det b^\dagger)^N\,|0\rangle\,.\ee
where $c_N$ is normalization coefficient and $N$ is the natural number that specifies the representation: $ \hat{N}\,|z_0\rangle \,=\,N\,|z_0\rangle$.
All other states in the representation space are obtained by the action of rising operators given
in (\ref{2.5a}): such states contain besides $nN$ $b$-particles a finite number of $ab$ pairs.\vskip 1cm

\section{Appendix C: The invariant Laplacian}
The invariant Laplacian operator $\Delta_N$ on the Bergman domain
is defined by: \begin{equation}
 T_g \Delta_N f(Z)=\Delta_N T_g f(Z)
\end{equation}
The interested could find in (\cite{Hua}) the explicit form of $\Delta_N$ for
arbitrary type I Cartan domain. For $D=SU(m,1)/U(m)$ it reads (\cite{Zh}\cite{PZh}):
\begin{equation}
 \Delta_N=(1-|Z|^2)(\sum_{i=1}^m\frac{\partial^2}{\partial \bar z_i \partial z_i}-\bar R R-N \bar R)
\end{equation}
where $R=\sum_1^m z_i\partial/\partial z_i$ is the first-order
differential operator. Here we consider only the radial part of the
invariant Laplacian and restrict ourselves to the discrete set of
eigenfunctions.

The eigenfunction of $\Delta_N$ reads (\cite{Zh}\cite{PZh}):
\begin{equation}
\phi_\lambda(Z)=(1-|Z|^2)^{(-N+2-i\lambda)/2}  F(\frac{N+2-i\lambda}{2},\  \frac{-N+2-i\lambda}{2};\ m;\ |Z|^2)
\end{equation}
where $F(\frac{N+2-i\lambda}{2},\  \frac{-N+2-i\lambda}{2};\ m;\ |Z|^2)$ is the hypergeometric function,  $\lambda$ is an complex number:
\begin{equation}
 \lambda=\alpha(\textbf{g}_c),
\end{equation}
where $\alpha$ are the roots for the complexified Lie algebra $su(m, 1)$ (see \cite{knapp}).
The eigenvalues read:
\begin{equation}
 -\frac{1}{4}((N-2)^2+\lambda^2).
\end{equation}

The discrete series of representation corresponds to the case
\begin{equation}
 \lambda=i(N-2-2l)
\end{equation}
where $N$ is a positive integer and $l=0, 1, \cdots, \big[\frac{N-2}{2}\big]$ we have discrete spectrum, which contains $l(l+2-N)$ points, $l=0, 1, \cdots ,\big[\frac{N-2}{2}\big]$ .

{\bf Acknowledgement }\  This paper is based on the previous joint work and interesting discussions
with Harald Grosse and Peter Presnajder. 
The author is also grateful to Vincent Rivasseau, Robin Zegers and Genkai Zhang for useful discussions. 
This work was partly supported by the ERC Starting Grant CoMBoS (grant agreement n.239694).

\end{document}